\documentclass[twocolumn,
amsmath,amssymb]{revtex4}

\usepackage{amssymb}
\usepackage[dvips]{graphicx}
\usepackage{url}
\begin{document}

\title{Controlling free superflow, dark matter and luminescence
rings of excitons in quantum well structures}.
\author{A. S. Alexandrov  and S. E. Savel'ev}

\affiliation{Department of Physics, Loughborough University,
Loughborough LE11 3TU, United Kingdom}
\maketitle

\textbf{Following the discovery of Bose-Einstein condensation (BEC) in ultra cold atoms [E. Gosta, Nobel Lectures in Physics (2001-2005), World Scientific (2008)],
there has been a huge experimental and theoretical
push to try and illuminate a superfluid
state of Wannier-Mott excitons. Excitons in quantum wells, generated by a laser pulse, typically diffuse only a few micrometers from the spot they are created.
However, Butov et al. [1] and Snoke et al. [2] reported luminescence from indirect and direct excitons hundreds of micrometers away from the laser excitation spot in double and single quantum well (QW) structures at low temperatures.  This luminescence appears as
a ring around the laser spot with the dark region
between the  spot and the ring. Developing
the theory of a free superflow of Bose-liquids we
show that the macroscopic luminesce rings and
the dark state observed in \cite{but,snoke} are signatures of the coherent superflow of condensed excitons at temperatures below their
Berezinskii-Kosterlitz-Thouless (BKT) transition
temperature \cite{bkt}. To further  verify the dark excitonic superflow we propose several keystone experiments, including interference of  superflows from two laser spots, vortex formation, scanning of moving dipole moments, and a giant increase of the luminescence distance  by applying one-dimensional confinement potential. These experiments combined with our theory  will open a new avenue for  creating and controlling superflow of coherent excitons on nanoscale.}

\begin{figure}
\begin{center}
\includegraphics[angle=-0,width=0.43\textwidth]{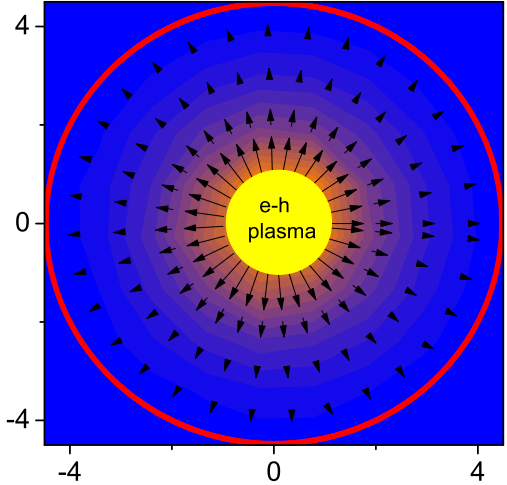}
\vskip -0.5mm \caption{(Color online) Superfluid of slowly decaying excitons  travels a macroscopic distance from the laser spot producing PL, when it stops far away from the photo-excited region. Relative length of vectors changes consistently with equation (\ref{j2}) for the exciton momentum. The momentum $k(\rho)$ drops to zero at the luminescence ring shown as a red ring. See also animation \cite{url1},
 where velocities in excitonic superfluid scale according to Eq.(\ref{j2}); excitons start flashing in the animation as soon as their momentum drops below $K_c$ and they emit photons.} \label{rings}
\end{center}
\end{figure}

The dramatic appearance of luminescence rings with radii of several
hundred microns in quantum well structures has been originally
attributed to a boundary between a positive hole gas diffusing from
the laser spot and a negative electron gas located well
outside the spot \cite{rapport, butov2}. This implies an energetically unfavorable separation of the
charges into hole-dominant positive and electron-rich negative regions on a macroscopic scale.
The charge separation  model  reproduced some basic features  of the luminescence ring formation. However,  there are several observations that are not accounted for by this simple model including  a  non-monotonic dependence of the ring
radius on laser power and significant  short-range Coulomb interactions of the carriers \cite{snoke2}. The model is also refutable on the
grounds of a rather short exciton formation time, typically  ten
picoseconds or less for  relevant densities ($ > 10^{10}$ cm$^{-2}$) of photo-carriers
\cite{damen,blom,time}. Indeed, photoelectrons
(mass $m$) almost immediately emit optical phonons with the frequency $\omega$, so that
their speed is capped at about  $v=\sqrt {2\hbar \omega/m} \approx 4\times 10^7$ cm/s.
Thus, the exciton formation mean-free path  turns out to be only a few
microns, which is too short  for  creating the charge separation on
the macroscopic scale of a few hundreds of micrometers far outside the
spot. This is consistent with  measurements of a  photoluminescence (PL)  spectrum in the magnetic field   which have shown a diamagnetic shift of the PL energy peak, indicating that the electrons and holes are bound
into thermalized pairs (i.e excitons)  in the region about a few tens of micrometers from the center of the spot \cite{shift}. As one gets closer to the
center of the spot, this shift
is changed from  blue to a red one indicating an
electron-hole plasma inside the spot, where electron-hole
pairs rapidly form hot excitons by emission of phonons. Therefore, excitons, which can condense in  a coherent superfluid, are formed from electron-hole plasma at  rather short distances from the center of the laser spot, Fig.(\ref{rings}).

 Importantly, since
 the
electron-hole plasma expands inside the spot due to the
 Coulomb repulsion  \cite{sav}, energy and momentum conservation requires that excitons initially form in states with  finite center-of-mass momenta, $K$, which do not  couple to  light directly. If the lattice temperature, T is below the
 Berezinskii-Kosterlitz-Thouless (BKT)  transition temperature, T$<$ T$_{KT}\approx \hbar^2
n_{ex}/k_Bm_{ex}$, which is about a few Kelvin for the relevant exciton densities, $n_{ex}\gtrsim 10^{10}$cm$^{-2}$,  part of excitons should  form  a  coherent  state with a certain value of their momentum,  $K=K_0$  taken from electrons and holes forming pairs. Thus one can expect  transitions of electrons and holes  from their plasma into
a flowing superfluid state of composed bosons in contrast to the more conventional stationary BEC of ultra cold atoms.
As a result we expect an excitonic radial
\emph{supercurrent} appearing at some  distance, $R_0$ inside or just around
the laser spot, $J_s(R_0)\equiv J_0  \neq 0$.
This new state of matter allows excitons to travel on a macroscopic distance from the laser spot producing PL far away from the photo-excited region (Fig. 1), as observed in Refs.\cite{but,snoke}.

\section*{Model}

To describe the concomitant free  superflow of two-dimensional (2D) Bose-liquid of excitons we apply  a generalized  Gross-Pitaevskii equation
\cite{gross} for the order-parameter, $\psi({\bf R})$ taking into
account the exciton  recombination rate $\gamma/\hbar$, as originally proposed by Keldysh \cite{keldysh}:
\begin{equation}
-\left({\hbar^2\over{2m_{ex}}} \Delta +\mu\right) \psi ({\bf R}) +V
|\psi ({\bf R})|^2\psi ({\bf R}) - i{\gamma \over{2}}\psi ({\bf
R})=0. \label{GP}
\end{equation}
Here  $\mu$ and $V$ conveniently parameterize the average
 superfluid density and the  short-range repulsion,
 respectively.

Introducing  dimensionless amplitude, $f({\bf r})=(V/\mu)^{1/2}|\psi
({\bf R})|$, and  the current density, ${\bf j}({\bf r})=f({\bf r})^2
\nabla \phi ({\bf R})$, Eq.(\ref{GP}),  is reduced to the following
two equations
\begin{equation}
- \Delta f-f+f^3+{j^2 \over{f^3}}=0, \label{GPreal}
\end{equation}
and
\begin{equation}
\nabla \cdot {\bf j}=-\beta f^2 \label{cont}.
\end{equation}
where $\phi({\bf R})$ is the  phase of the order
 parameter and   coordinates ${\bf r}= {\bf R}/\xi$ are  normalized on the coherence length $\xi=(\hbar^2/2m_{ex}\mu)^{1/2}$. Eq.(\ref{GPreal}) is the familiar Ginzburg-Landau equation in
the presence of the current and Eq.(\ref{cont}) is the continuity
equation with the dimensionless decay rate,
$\beta=\gamma/2\mu$.

These equations are grossly simplified in the case of a low decay rate, $\beta \ll 1$. In this case the amplitude
of the order parameter and the current density  change at the length
scale of the order of $1/\beta \gg 1$ much larger than the coherence
length, so that the first term ("kinetic pressure") in
Eq.(\ref{GPreal}) is negligible. Then the current density is
expressed via the amplitude as
\begin{equation}
j \approx f^2 \sqrt{1-f^2}. \label{j}
\end{equation}
There are two brunches described with Eq.(\ref{j}),  but only one, which counter-intuitively corresponds to a larger order parameter $f^2\geqslant 2/3$ and smaller momentum $k=\sqrt{1-f^2}\leqslant 1/\sqrt{3}$, is stable for small enough $\beta$, Fig.\ref{brunch} (see supplementary material).
\begin{figure}
\begin{center}
\includegraphics[angle=-0,width=0.43\textwidth]{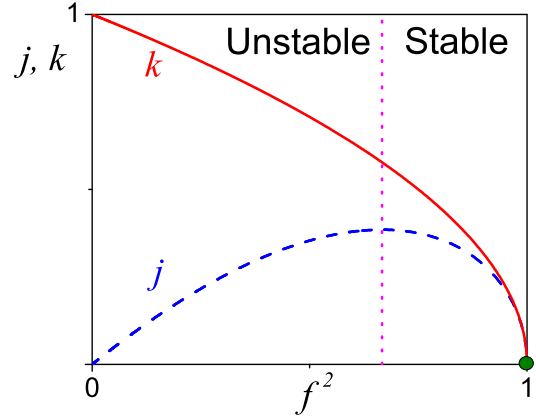}
\vskip -0.5mm \caption{(Color online) Supercurrent density and exciton momentum  as a function of the order parameter. At relatively small exciton recombination rates the stable uniform superflow (to the right from the vertical dotted line)  is  realized for  lower momenta, $k < 1/\sqrt{3}$ and higher order parameters, $f^2 > 2/3$. The dot corresponds to the luminescence region, where excitons "burn" radiating light. Note that the order parameter (i.e. the  superfluid density) \emph{increases}  approaching the luminescence region.} \label{brunch}
\end{center}
\end{figure}

\vspace{0.5cm}
\emph{A. One-dimensional (1D) superflow and PL.}

Let us first discuss the stable superflow
in 1D structures which can be made by applying a 1D confinement potential to experimentally available 2D samples. By using the boundary condition, $j(0)= V(m_{ex}/2\mu^3)^{1/2} J_0$, and assuming that $R_0/\xi \ll 1/\beta$, Eq. (\ref{cont}), written in the form $dk/dx=\beta
(k^2-1)/(1-3k^2)$ for the exciton momentum $k(x)$,
can be easily analytically solved:
\begin{equation}
3(k_0-k) +\ln \left[{(1+k)(1-k_0)\over{(1-k)(1+k_0)}}\right]= \beta
x, \label{k}
\end{equation}
where  $k_0$ is found from $k_0 -k_0^3=j(0)$.

To determine location of luminescence, we use that  excitons can  radiate in 1D and 2D by resonant emission of photons only if their momentum is inside the photon cone \cite{cone,decay}, namely if superfluid center-of-mass momentum is small enough:
  $K \leqslant K_c = E_g\sqrt{\epsilon}/c$.  Here $E_g$ is the semiconductor band gap
and $\epsilon$ is the dielectric constant.  The exciton PL intensity is
determined by the fraction of excitons inside the cone, where their
decay rate $\beta$ is strongly enhanced  \cite{decay}.
For a narrow photon
cone,  $K_c\ll K_0$,  the position  of a bright luminescence stripe, $x_{ring}$ is found by
taking $k=0$ in Eq.(\ref{k}),
\begin{equation}
x_{ring}= \beta^{-1}\left[3k_0-\ln \left({1+k_0\over{1-k_0}}\right)
\right]. \label{x}
\end{equation}
Since $k_0$ is of the order of one, the distance $x_{ring}$ is macroscopic, i.e. large compared
with the coherence length, as  $1/\beta \gg 1$.

\vspace{0.5cm}
\emph{B. Controlling PL in 2D  superflow.}

\begin{figure}
\begin{center}
\includegraphics[angle=-0,width=0.52\textwidth]{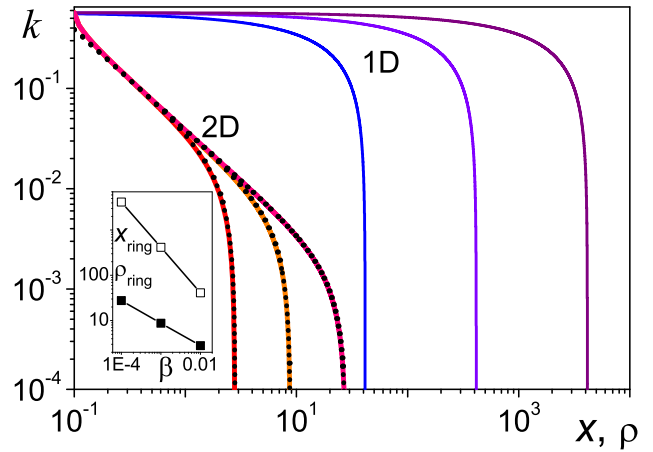}
\vskip -0.5mm \caption{(Color online) The  exciton momentum  as a function of the distance from the laser spot obtained using Eqs. (\ref{k},\ref{j2}) for three different values of the decay rate $\beta=10^{-2},\ 10^{-3},\ 10^{-4}$ in both 1D and 2D geometries.  Dotted black lines correspond to  solutions of the linearized 2D continuity equation. They practically coincide with the numerical solutions of the exact equation except of very short distances form the laser spot. Inset shows the dependence of the position of the luminescence region for these decay rates in log-log scale. It is seen that $x_{ring}$ scales as $1/\beta$, while $\rho_{ring}$ scales as $1/\sqrt{\beta}$. This results in $x_{ring}\gg\rho_{ring}$, that is luminescence stripes formed in 1D structures  locate much farther from the excitation spot comparing with  luminescence rings  in 2D geometry. Gradually changing the geometry from 2D to 1D  should move the luminescence  farther  from the laser spot.} \label{2Dprofile}
\end{center}
\end{figure}

Let us now consider an already implemented experimental situation \cite{but,snoke}, where excitons created around the central laser spot propagate freely in a plane.
In contrast to  the original proposal of Ref.\cite{but} we suggest
 a continuous \emph{super}flow of excitons out of the excitation spot  under steady-state photoexcitation, rather than their normal  drift and diffusion. For isotropic 2D superflow the order parameter depends only on the distance $\rho$ from the spot. Then 2D continuity equation (\ref{cont}) (written for the exciton momentum $k(\rho)=\sqrt{1-f^2(\rho)}$) takes a simple one-dimensional form,
\begin{equation}
{dk\over{d\rho}}+{k-k^3\over{\rho(1-3k^2)}}=\beta
{k^2-1\over{1-3k^2}}, \label{j2}
\end{equation}
 The stable numerical  solution of this equation is shown in Fig. \ref{2Dprofile} together with  $k(x)$ dependence of the momentum in 1D structures using Eq.(\ref{k}). The analytical solution of  linearized equation (\ref{j2}), $k'_\rho+k/\rho=-\beta$, which is $k(\rho)=k_0\rho_0/\rho -\beta \rho/2$ practically coincides  with  the numerical solution of the nonlinear Eq.(\ref{j2}) in the whole relevant space
 ($\rho_0$ is the distance from the center of the spot where the superflow starts).
As in the 1D case, PL ring radius corresponds to $k(\rho_{ring})=0$, so that $\rho_{ring}=(2k_0\rho_0/\beta)^{1/2}.$ However, different from 1D geometry, where $x_{ring}$ scales as $1/\beta$, the ring radius scales as $\rho_{ring}\propto 1/\sqrt{\beta}$ in the 2D case (see inset in Fig.(\ref{2Dprofile})).  Therefore, the radius of PL for 2D case occurs to be much shorter than a distance between the laser spot and 1D  luminescence stripe. This prediction can be verified by channeling the superfluid, that is by applying 1D confinement potential  in order to controllably transform 2D to 1D geometry and  check if PL shifts away from the spot.

In dimensional units the PL ring radius is
$R_{ring}\approx  \sqrt{2\hbar  R_0 K_0/(m_{ex}\gamma)}$.
The  distance from the center of the spot, $R_0$,  where the superflow starts with the momentum $K_0$, can be  estimated using the energy
conservation. The average electron potential energy inside the spot is $Vn$, where  $V \approx 2\pi \alpha d $, $d$ is
the effective separation between the electron and hole layers ($\approx 10$ nm),  $\alpha= e^2/(4\pi \epsilon_0\epsilon)\approx 2.3\times 10^{-28}/\epsilon$ Jm, and $\epsilon$ is the static dielectric constant of the semiconductor. It transforms into the exciton kinetic energy, i.e. $Vn=K^2/2m_{ex}$,
so that $K_0\approx \sqrt{4\pi \alpha d n m_{ex}}$.  The radius $R_{0}$,  where the superfluid is formed is about $R_{0}=v\tau_{ex}$, where
$ \tau_{ex}$ is the exciton formation time. As a result we find
\begin{equation}
R_{ring} \approx \sqrt {\hbar v \tau_{ex} (16\pi d\alpha n/m_{ex})^{1/2}\over{\gamma}} \label{ring}.
\end{equation}
Importantly  $\gamma$ in this estimate is the \emph{nonradiative} recombination rate since
the coherent excitons have their momentum outside the photon cone. The nonradiative lifetime, $\tau_0=\hbar/\gamma$ is an order of magnitude or more   longer than  the  exciton radiative lifetime \cite{nonrad}.  Remarkably, using the realistic material parameters
$\tau_{ex}=1$ps \cite{time}, $\tau_0=10$ $\mu$s \cite{nonrad}, $m_{ex}= 0.2 m_e$, $\epsilon = 13$ in Eq.(\ref{ring}) yields  $R_{ring}$ about $500$ $\mu$m for the photoelectron density $n= 10^{10}$ cm$^{-2}$, explaining the macroscopic radius of the rings. Dissipation processes  \emph{stabilize} the steady-state superflow of coherent excitons (see supplementary material), in contrast with  a normal state flow, where individual excitons can scatter to a lower energy state emitting acoustic phonons.

Also the dependence of the ring size and its PL intensity, $I$, on the excitation power, $P$, can be readily understood in the framework of the free superflow. According to Eq.(\ref{ring}) the ring radius scales as  $R_{ring} \propto n^{1/4}\tau_{ex}^{1/2}$.
The ring luminescence intensity, $I$ is  proportional to the exciton density on the ring, $n_{ex}$.  Excitons are pumped into the coherent state at $R_0$ with the rate $n/\tau_{ex}$, where the exciton formation time, $\tau_{ex}$ is inverse proportional to the photocarrier density $n$ and it strongly depends on the exciton momentum $K_0$ \cite{time}.  After flowing the distance $R_{ring}$ excitons decay radiating light, so their density on the ring scales with the photocarrier density as
$n_{ex} \propto R_0 n/R_{ring}\tau_{ex}$.  Parameterizing the momentum  dependence of the exciton formation time
as $\tau_{ex} \propto K_0^r/n$
  yields the following  scaling: $R_{ring} \propto n^{(r-1)/4}$ and
$I \propto n^{(5-r)/4}$. Hence,
 we find  $R_{ring} \propto P^{(r-1)/4}$,
 $I \propto P^{(5-r)/4}$ in the case of a linear photoelectron population ($n \propto P$), and $R_{ring} \propto P^{(r-1)/8}$,
 $I \propto P^{(5-r)/8}$ in the case of  $n\propto P^{1/2}$. Since  $r$ can be a large number ($\geqslant 5$) \cite{time}, our theory predicts an increase of the ring radius and a decrease of the PL intensity  with higher excitation power. In particular, the product $R_{ring} I$  is  proportional to $n$ for any exponent $r$   independent of modeling the exciton-formation-time, thus resulting in $R_{ring} I \propto P^{1/2}$ as observed \cite{but} for $n\propto P^{1/2}$.    More generally the dependence of $\tau_{ex}$ on the exciton momentum is essentially non-monotonous \cite{time}, so that the ring radius and the PL intensity  may depend on the excitation power in a more complicated fashion as also observed \cite{snoke2}.

 There is a threshold density of photoelectrons $n_c$ in our theory in agreement with  the experiments \cite{but,snoke},  and hence the laser power below which the ring does not form, which can be found from $K_0=K_c$. At this and lower excitation power coherent exciton superfluid radiates just at the point of its formation inside or close to the laser spot. Using our estimate of $K_0$ one obtains $n_c = E_g^2\epsilon /(4\pi c^2\alpha d m_{ex})\approx 0.2 \times 10^{10}$ cm$^{-2}$.

While a detailed description of the outer ring
fragmentation into a periodic array observed in Ref. \cite{but} is outside the scope of the current paper,  the
periodic patterns  are anticipated in our theory since  the uniform coherent state of the outer ring is unstable due to a strongly enhanced recombination rate inside the photon cone (see supplementary material).

\vspace{0.5cm}
\emph{C. Two-spot superflow and dark matter.}

Experiments with two rings created by spatially separated
laser spots  revealed that
the rings attract one another, deform, and then open
towards each other \cite{butov2}. This happens
before the rings coalesce into a common oval-shaped ring, suggesting the existence of a ``dark matter"
outside the rings that mediates the interaction. Such a ring attraction is hard to consistently interpret in the classical electron-hole plasma model \cite{sav}, where outer electron-rich regions from two spots naturally should repel each other. Here we show that this dark matter is the coherent superflow of excitons from two laser sports interfering with each other. Our  continuity equation (\ref{cont}) becomes now an equation for the two components of  the exciton momentum ${\bf k}=(k_x,k_y)$,
\begin{equation}
{\partial (1-k^2)k_x\over{\partial x}}+{\partial (1-k^2)k_y\over{\partial y}} =\beta
(k^2-1), \label{j22}
\end{equation}
 Using ${\bf k}=\nabla \phi$ it can be reduced to a nonlinear partial differential equation for the  phase of the order
 parameter $\phi(x,y)$  with the boundary condition $k=k_0$ on two small circles inside or just around every spot.
As numerically shown above, the linearized continuity equation describes well the superflow from a single spot on the relevant distance, Fig.(\ref{2Dprofile}), so that  we consider a linearized version of Eq.(\ref{j22}):
\begin{equation}
\Delta_2 \phi(x,y)=-\beta, \label{j4}
\end{equation}
where $\Delta_2=\partial^2/\partial x^2 +\partial^2/\partial y^2$. The individual ring size is about $ \sqrt{k_0\rho_0/\beta}$, so that  one  expects two independent PL rings when  the distance $d$ between the spots is large $d\gg\sqrt{k_0\rho_0/\beta}$. The interaction should appear at the distance comparable to the ring size.  If we assume that the superflow starts at a  small distance from the centers of the spots such that $\rho_0  \ll k_0/\beta$, then a solution of Eq.(\ref{j4})
\begin{equation}
\phi(x,y)=-\beta {x^2+y^2\over{4}} +k_0\rho_0 \ln (\rho_1 \rho_2) \label{solution}
\end{equation}
approximately satisfies the boundary conditions for $d/2 \lesssim \sqrt{k_0\rho_0/\beta}$. Here $\rho_{1,2} = \sqrt{(x\pm d/2)^2+y^2}$ is the distance from each spot, respectively. If we further assume that $\rho_0 \lesssim  k_c^2/(\beta k_0) $, then the condition for  PL, $k\leqslant k_c$, is satisfied at the contour found from
\begin{equation}
{1\over{(x-d/2)^2+y^2}}+{1\over{(x+d/2)^2+y^2}}={\beta\over{2k_0\rho_0}},\label{contour}
\end{equation}
and shown in Fig.\ref{spots}. The series of the obtained contours agree well with the experimental observations \cite{butov2}:
\begin{figure}
\begin{center}
\includegraphics[angle=-0,width=0.40\textwidth]{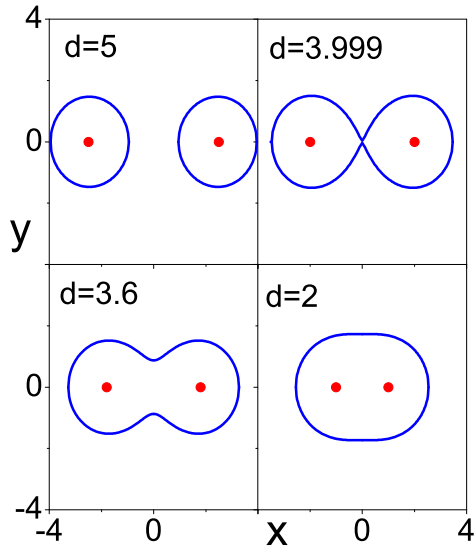}
\vskip -0.5mm \caption{(Color online) PL contours from two laser spots for different distances $d$ between the spots (measured in  $\xi \sqrt{k_0\rho_0/\beta}$) obtained by using Eq.(\ref{contour}). As in the experiment \cite{ butov2} the rings first attract one another and, then, open towards each other. See also animations,  where PL patterns from two laser spots separated by a fixed distance evolve  with increasing  $k_0\rho_0/\beta$ from 0.01 to 0.18 \cite{url2}.} \label{spots}
\end{center}
\end{figure}
as the spots are brought closer,
the rings deform, and then open
towards each other. This happens well
before the rings coalesce into a common oval-shaped ring (right lower panel in Fig.(\ref{spots})).

Apart from the normal state  diffusion in the charge-separation model, an alternative ``superfluid"  interpretation  of  PL rings    in
double and single QWs has been hampered by the lack of the theory of the free 2D superflow.
Based on our theory, we suggest that the macroscopic luminescence
rings and the dark matter in QWs  originate  in the   slowly decaying superflow
of  excitons out of the laser spot at temperatures below their
BKT  temperature. The existence of a critical temperature about a few Kelvin of the macroscopic  ring onset \cite{butov2} strongly supports our conclusion. Our results imply that Butov et al. \cite{but} and Snoke et al. \cite{snoke} discovered a new state of matter: the exciton superflow  in the dark
 state. This state can be further  experimentally verified with light scattering, STM or any other spectroscopy sensitive to the moving dipole moments. Another piece of evidence for  the superflow should be an observation of a quantum interference and/or  vortex structures with the characteristic coherence length $\xi =\hbar (2m_{ex}n_{ex}V)^{-1/2}$ of about $15$ nm. Last but not least prediction of the theory
is a giant increase of the distance to the luminescence region by applying a 1D confinement potential tuning the geometry  from 2D to 1D.

\subsection*{Acknowledgements}
We thank J. Beanland, L. V. Butov,  V. V. Kabanov,  M. Yu. Kagan, and V. A. Khodel for illuminating comments and discussions.
This work has been partially supported by the Royal Society and the Leverhulme Trust (UK).


\begin{thebibliography}{90}
\bibitem{but} Butov, L. V., Gossard, A. C. \& Chemla, D. S. Macroscopically
ordered state in an exciton system. \emph{Nature} \textbf{418},
751�754 (2002).
\bibitem{snoke}
Snoke, D., Denev, S., Liu, Y., Pfeiffer, L. \& West, K. Long-range
transport in excitonic dark states in coupled quantum wells.
\emph{Nature} \textbf{418}, 754�757 (2002).
\bibitem{bkt} Berezinskii, V. L. Destruction of long-range order in one-dimensional and two-dimensional systems possessing a continuous symmetry group. II. Quantum systems. \emph{Sov. Phys. JETP} \textbf{34}, 610�616
(1972); Kosterlitz, J. M. \& Thouless, D. J. Ordering, metastability
and phase-transitions in 2 dimensional systems. \emph{J. Phys. C}
\textbf{6}, 1181�1203 (1973).
\bibitem{rapport} Rapaport, R., Gang Chen, G.,  Snoke, D., Simon, S. H.,  Pfeiffer, L.,  West, K.,   Liu, Y. \&
Denev, S. Charge Separation of Dense Two-Dimensional Electron-Hole
Gases: Mechanism for Exciton Ring Pattern Formation. \emph{Phys.
Rev. Lett.} \textbf{92},  117405 (2004).
\bibitem{butov2}  Butov, L. V.,  Levitov, L. S.,  Mintsev, A. V.,   Simons, B. D.,  Gossard, A. S.,  \& Chemla1, D. S. Formation Mechanism and Low-Temperature Instability of Exciton
Rings. \emph{Phys. Rev. Lett.} \textbf{92}, 117404 (2004).
\bibitem{snoke2} Snoke, D., Denev, S., Liu, Y., Simon, S., Rapaport, R., Chen, G., Pfeiffer, L., \& West, K.  Moving beyond a simple model of luminescence rings in quantum well
structures. \emph{J. Phys.: Condens. Matter} \textbf{16},
S3621-S3627 (2004).
\bibitem{damen} Damen, T. C.,  Shah, J.,   Oberli, D. Y.,  Chemla, D. S.,  Cunningham, J. E., \&  Kuo, J. M. Dynamics of exciton formation and relaxation in GaAs quantum wells. \emph{Phys. Rev. B} \textbf{42}, 7434�7438 (1990).
\bibitem{blom} Blom, P. W. M.,   van Hall, P. J.,  Smit, C.,  Cuypers, J. P., \&
Wolter, J.H. Selective exciton formation in thin GaAs/AlxGa1-xAs quantum wells
\emph{Phys. Rev. Lett.} \textbf{71}, 3878 (1993).
\bibitem{time}  Oh, I. K.,  Singh, J.,  Thilagam, A.  \& Vengurlekar, A. S. Exciton formation assisted by LO phonons in quantum
wells. \emph{Phys. Rev. B }\textbf{62}, 2045 (2000).
\bibitem{shift} Stern, M.,  Garmider, V.,  Segre, E.,  Rappaport, M.,  Umansky, V., Levinson, Y. \&  Bar-Joseph I. \emph{Phys. Rev. Lett.} \textbf{101}, 257402 (2008).
\bibitem{sav}  Paraskevov, A. V.,  \&  Savel�ev, S. E. Ring-shaped luminescence patterns in a locally photoexcited electron-hole bilayer. \emph{Phys. Rev. B} \textbf{81}, 193403 (2010).
\bibitem{url1} \url{http://www-staff.lboro.ac.uk/~phss/super fluid 1.swf}
\bibitem{gross}
 Gross, E. P. Structure of a quantized vortex in boson systems. \emph{Nuovo Cimento} ${\bf 20}$, 454 (1961);  Pitaevskii, L.P.
Vortex Lines in an Imperfect Bose Gas. \emph{Zh. Eksp. Teor. Fiz.} {\bf 40}, 646 (1961) (\emph{Soviet Phys. JETP} ${\bf
13}$, 451 (1961)).
\bibitem{keldysh}  Keldysh, L. V. Problems of Theoretical Physics
 (Nauka, Moscow, 1972) p. 433.
\bibitem{cone} Feldmann, J., Peter, G., G�obel, E.O., Dawson, P., Moore,
K., Foxon, C., and Elliott, R.J. Linewidth dependence of
radiative exciton lifetimes in quantum wells. \emph{Phys. Rev.
Lett.} 59, 2337 (1987).
\bibitem{decay}  Andreani, L. C. Radiative lifetime of free excitons in quantum wells. \emph{Solid State Commun.}  \textbf{77},  641-645 (1991).
\bibitem{nonrad}  Starukhin, A. N., Nelson, D. K., Yakunenkov,  A. S.,  \&  Razbirin, B. S. Radiative and nonradiative recombination
kinetics of indirect bound excitons studied
by time-resolved level anticrossing
experiments.\emph{ J. Phys.: Condens. Matter} \textbf{20 }, 055208 (2008).
\bibitem{url2} \url{http://www-staff.lboro.ac.uk/~phss/Rings.swf}
\end{thebibliography}
\end{document}


\title{Controlling free superflow, dark matter and luminescence
rings of excitons in quantum well structures.

Supplementary material: instability analysis}.
\author{A. S. Alexandrov and S. E. Savel'ev}

\affiliation{Department of Physics, Loughborough University,
Loughborough LE11 3TU, United Kingdom}
\maketitle

Instabilities of the stationary  superflow   can be analyzed
 using a
generalised time-dependent Gross-Pitaevskii equation,
 written in our units as
\begin{equation}
-(\Delta+1) \psi ({\bf r},t)+\psi^2({\bf r},t) \psi^*({\bf r},t)-i\beta \psi ({\bf r},t)=(i-u) {\partial
\psi ({\bf r},t)\over{\partial t}}. \label{sGP}
\end{equation}
Here $u$ accounts for any kind of  dissipation, for instance  emission of phonons as in the
time-dependent Ginzburg-Landau equation \cite{kopnin}.

One can linearize this equation with respect to small fluctuations
of the uniform superlow  $\psi_s({\bf r})=f({\bf r}) \exp[i \phi({\bf r}]$ satisfying the stationary GP equation.  
Taking  $\psi(x,t)= \psi_s(x) +\tilde{\psi}(x,t)$ and $\tilde{\psi}(x,t)\equiv \eta(x,t)\psi_s(x)$ yields for small $|\eta(x,t)|\ll 1$
\begin{equation}
-\eta''_{xx}-2\eta'_{x}(f^{-1}f'_{x}+i\phi'_x)+f^2(\eta+\eta^*)=(i-u)\dot{\eta} \label{ssGP}
\end{equation}
in 1D geometry.
The stationary amplitude $f(x)$ and the phase $\phi(x)$ satisfy the following pair of equations:
\begin{eqnarray}
&&-f''_{xx}+f(\phi'_x)^2-f+f^3=0 \cr
&&f\phi''_{xx}+2f'_x\phi'_x=-\beta f,
\end{eqnarray}
where the second equation is the continuity equation. At small $\beta \ll 1$ one can neglect the second derivative of the stationary amplitude  $f''_{xx}\propto \beta^2$,  so that $\phi'_x=(1-f^2)^{1/2}\equiv k$ and $f'_x=-\beta fk/(3k^2-1)$, which substitution into Eq.(\ref{ssGP}) yields
 \begin{equation}
-\eta''_{xx}-2\eta'_{x}\left(ik-{\beta k\over{3k^2-1}}\right)+(1-k^2)(\eta+\eta^*)=(i-u)\dot{\eta} \label{sssGP}
\end{equation}
\begin{figure}
\begin{center}
\includegraphics[angle=-0,width=0.35\textwidth]{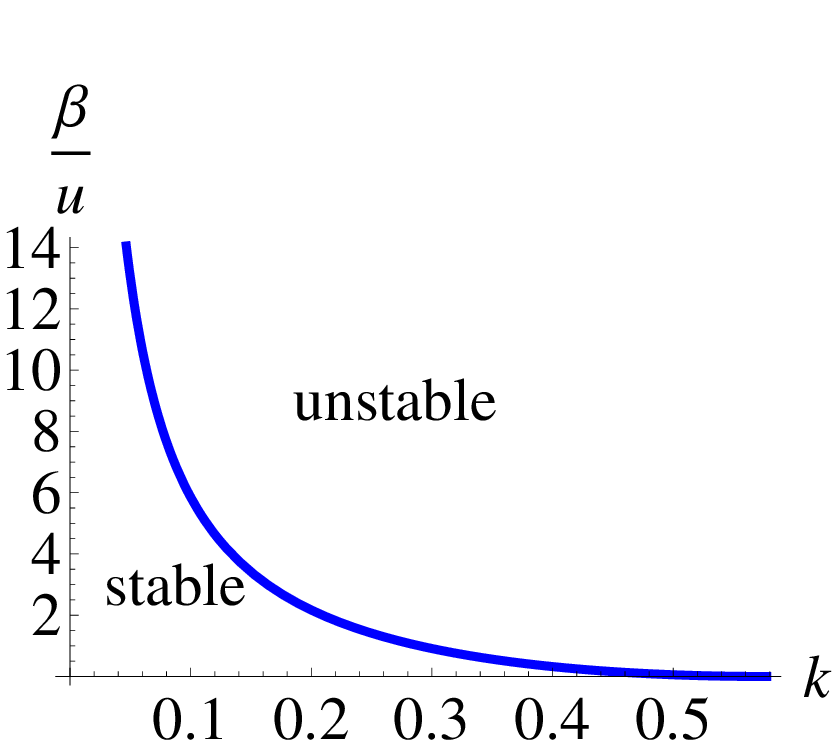}
\vskip -0.5mm \caption{Domains of stable and unstable uniform superflow. Increasing the exciton decay rate $\beta$ relative to the dissipation rate $u$ shrinks the stability region to lower values of the coherent momentum $k$.}
\end{center}
\end{figure}
For a sufficiently long time, $t\lesssim 1/\beta \gg 1$, a solution can be found in the form:
$\eta(x,t)=a \exp[i(\omega t-qx)]+b \exp[-i(\omega^*
t-qx)]$, where coefficients $a$ and $b$ satisfy the generalized Bogoliubov
equations,
\begin{eqnarray}
&&(1-k^2-q^2-2qk-{2i\beta qk\over{3k^2-1}}+\omega+iu\omega)a=(k^2-1)b,\cr
 && (1-k^2-q^2+2qk-{2i\beta qk\over{3k^2-1}}-\omega+iu\omega)b=(k^2-1)a, \nonumber
\end{eqnarray}
with the eigenvalues,
\begin{equation}
\omega=2kq \pm \sqrt{(1-k^2+q^2-{2i\beta qk\over{3k^2-1}}+iu\omega)^2-(1-k^2)^2}.
\end{equation}
In the absence of any decay and dissipation we recover the familiar Bogoliubov
spectrum of excitations, shifted by the current,
\begin{equation}
\omega^{\pm}_B=2kq \pm \sqrt{q^4+2(1-k^2)q^2}.
\end{equation}
The decay $\beta$ and the  dissipation $u$ lead to an imaginary part of $\omega$, which is
\begin{equation}
\Im{\omega}= \pm
\left(u\omega^{\pm}_B-{2\beta kq\over{3k^2-1}}\right){(1-k^2+q^2)\over{\sqrt{q^4+2(1-k^2)q^2}}}
\end{equation}
for small $u,\beta \ll 1$. The instability appears when $\Im{\omega}$ is negative. In the absence of  the decay ($\beta=0$) ,  this condition
corresponds to the negative $\omega^{+}_B$, or the positive
$\omega^{-}_B$, which is the case if
\begin{equation}
4k^2 \geqslant q^2+2(1-k^2).
\end{equation}
Hence the uniform superflow with sufficiently  small momentum, $k \leqslant
1/\sqrt{3}$ is stable, while the higher momentum states with $k >
1/\sqrt{3}$ are unstable. A finite decay $\beta \neq 0$ shrinks the region of the uniform superflow to $k < 1/\sqrt{3}$ as shown in Fig.1.